\catcode`\@=11

\@addtoreset{equation}{section}

\def\@normalsize{\@setsize\normalsize{15pt}\xiipt\@xiipt
\abovedisplayskip 14pt plus3pt minus3pt%
\belowdisplayskip \abovedisplayskip
\abovedisplayshortskip  \z@ plus3pt%
\belowdisplayshortskip  7pt plus3.5pt minus0pt}

\def\small{\@setsize\small{13.6pt}\xipt\@xipt
\abovedisplayskip 13pt plus3pt minus3pt%
\belowdisplayskip \abovedisplayskip
\abovedisplayshortskip  \z@ plus3pt%
\belowdisplayshortskip  7pt plus3.5pt minus0pt

\def\@listi{\parsep 4.5pt plus 2pt minus 1pt
	    \itemsep \parsep
	    \topsep 9pt plus 3pt minus 3pt}}

\def\underline#1{\relax\ifmmode\@@underline#1\else
	$\@@underline{\hbox{#1}}$\relax\fi}
\@twosidetrue

\catcode`@=12

\catcode`\@=11

\catcode`\@=12

\relax

\def\figcap{\section*{Figure Captions\markboth
	{FIGURECAPTIONS}{FIGURECAPTIONS}}\list
	{Fig. \arabic{enumi}:\hfill}{\settowidth\labelwidth{Fig. 999:}
	\leftmargin\labelwidth
	\advance\leftmargin\labelsep\usecounter{enumi}}}
 \relax
\def\tablecap{\section*{Table Captions\markboth
	{TABLECAPTIONS}{TABLECAPTIONS}}\list
	{Table \arabic{enumi}:\hfill}{\settowidth\labelwidth{Table
	999:}
	\leftmargin\labelwidth
	\advance\leftmargin\labelsep\usecounter{enumi}}}
 \relax
\def\reflist{\subsubsection*{References\markboth
	{REFLIST}{REFLIST}}\list
	{[\arabic{enumi}]\hfill}{\settowidth\labelwidth{[999]}
	\leftmargin\labelwidth
	\advance\leftmargin\labelsep\usecounter{enumi}}}
 \relax

\catcode`\@=11

\def\FERMIPUB{}

\def\ps@headings{\def\@oddfoot{}\def\@evenfoot{}
\def\@oddhead{\hbox{}\hfill
	\makebox[.5\textwidth]{\raggedright\ignorespaces --\thepage{}--
	\hfill {\rm FERMILAB--Pub--\FERMIPUB}}}
\def\@evenhead{\@oddhead}
\def\subsectionmark##1{\markboth{##1}{}}
}

\ps@headings

\relax

\newskip\humongous \humongous=0pt plus 1000pt minus 1000pt

\newif\ifdtup

\def\beq{\begin{equation}}
\def\eeq{\end{equation}}

\def\beqn{\begin{eqnarray}}
\def\eeqn{\end{eqnarray}}
\relax

\def\dotx{\dotx{\dot\overline{x}}}

\documentstyle[art12,A4]{article}
\def\today{\number\day\space
     \ifcase\month\or
       January\or February\or March\or April\or May\or June\or
       July\or August\or September\or October\or November\or
       December\fi
     \space\number\year}

\begin{document}
\begin{titlepage}
\begin{flushright}
CERN-TH.7191/94\\
Zurich University ZU-TH 8/94\\
\end{flushright}
\vfill
\begin{center}
{\large\bf ON THE MASS OF THE DARK COMPACT OBJECTS \\
 IN THE GALACTIC DISK}
\vfill
{\bf Ph.~Jetzer*}\\
\vskip 1.0cm
Theory Division, CERN\\
CH-1211 Geneva 23, Switzerland\\
and\\
Institute of Theoretical Physics, University of Zurich,
Winterthurerstrasse 190,\\
CH-8057 Zurich, Switzerland\\
\end{center}
\vfill
\begin{center}
Abstract
\end{center}
\begin{quote}
Recently the Polish-American collaboration OGLE
has reported the observation of
four  possible microlensing events by monitoring, over several months,
the brightness of millions of stars in the region of the
galactic bulge.
If these events are due to microlensing, the most accurate way
to get information on the mass of the dark
compact objects, that acted
as gravitational lenses, is to use the method of the mass moments.
Here I apply this method to the analysis of the events detected by
OGLE.
The average mass
turns out to be
0.28$M_{\odot}$, suggesting that the lens objects are faint disk
stars.
The same method applied to the five microlensing events detected so
far by the EROS and MACHO collaborations, which monitor
stars in the Large Magellanic Cloud,
leads to an average value of 0.08$M_{\odot}$ for
the dark compact halo objects.

\end{quote}
\vfill
\begin{flushleft}
CERN-TH.7191/94\\
March 1994\\
\vskip 0.5cm
$^*$ Supported by the Swiss National Science Foundation.
\end{flushleft}
\end{titlepage}
\newpage

\noindent{\bf 1. Introduction}\\

An important problem
in astrophysics is the nature of the non-luminous
matter in which galaxies are embedded. Its
presence is inferred from the
shape of the measured rotation curves. It is well possible that
the dark
matter in the halo and in the disk of the galaxies is made of
``brown dwarfs'' or Jupiter-like bodies,
which are aggregates of the primordial elements: hydrogen and helium,
and have masses in the range ${\rm O}(10^{-7}) < M/M_{\odot} <
{\rm O} (10^{-1})$  (De R\'ujula et al. 1992).
Paczy\'nski suggested a way to detect such objects in the halo and
in the disk of our
own galaxy using the gravitational lens effect (Paczy\'nski 1986,
1991; Griest et al. 1991).

Recently the French collaboration EROS (Aubourg et al. 1993)
and the American--Australian
collaboration MACHO (Alcock et al. 1993)
reported the possible detection of
altogether five microlensing events,
discovered by monitoring over several
years millions of stars in the Large Magellanic Cloud (LMC),
whereas the
Polish-American collaboration OGLE (Udalski et al. 1993)
has found four
microlensing events by monitoring stars located in the galactic bulge.

If these observations are true microlensing events, an
important question is
to determine the mass of the dark compact objects, that
acted as gravitational lenses. This knowledge
is also relevant
in order to infer the amount of dark matter in our galaxy and in its
halo.
The most appropriate way to compute the average mass is to use
the method of mass moments developed by De R\'ujula et al. (1991).

Here I apply this method to the four events detected by OGLE  by
looking at the galactic bulge and to the ones of EROS and
MACHO in the LMC. I also
compute for each event the value of the most likely mass of the lens
object.
Of course the very small number of observations at our disposal does
not yet
allow a precise determination of the mass distribution, so that
the present results are preliminary.
Nevertheless, it shows how in practice it will
be possible to get precise information, as soon as
a sufficient number of microlensing events will be available, on the
mass
distribution as well
as on what fraction they contribute to the total
dark mass in the halo or in the disk of our galaxy.
In the   following I will use the formulas and the notation derived
in De R\'ujula et al. (1991), to which I refer for more details on the
derivation; I refer to the paper by Jetzer \& Mass\'o (1994) as well,
where the first three discovered events in the LMC are discussed.\\

\noindent{\bf 2. Microlensing rates}\\

First I compute the microlensing rate $\Gamma$
for an experiment monitoring
stars in the galactic bulge in Baade's window of galactic
coordinates (longitude and latitude): $l = 1^{\circ}$, $b =
-3.^{\circ}9$.

Let $d$ be the distance of the massive halo or disk
object (MHO) to the line of
sight between the observer and a star in the galactic bulge,
$t = 0$ the instant
of closest approach, and $v_T$ the MHO velocity in the transverse
plane. The magnification $A$ as a function of time is calculated using
simple geometry, and is given by
\begin{equation}
A(t)=A[u(t)]=\frac{u^2+2}{u(u^2+4)^{1/2}}~, \label{eqno:1}
\end{equation}
where
\begin{equation}
u^2=\frac{d^2+v_T^2 t^2}{R_E^2}~. \label{eqno:2}
\end{equation}
The light curve is a universal function determined by the two
parameters $d/R_E$ and $v_T/R_E$. $R_E$ is the Einstein radius,
which is
\begin{equation}
R_E^2=\frac{4GMD}{c^2}x(1-x)=r_E^2 \mu x(1-x)~,  \label{eqno:3}
\end{equation}
with $M$ (respectively $\mu$) the MHO mass (in $M_{\odot}$ solar
mass units)
and $D~ (xD)$ the distance
from the observer to the source (to the MHO); $D = 8.5$ kpc is
the distance
to stars in the galactic bulge, and $r_E=1.25 \times 10^9$ km.
I use here the definition: $T=R_E/v_T$ (this is slightly
different from the definition used in De R\'ujula et al. (1991)).

The number density of disk stars per unit mass is given by
(Bahcall \& Soneira 1980)
\begin{equation}
\frac{dn}{dM}=\frac{dn_0}{dM}~
\exp\left(-\frac{D x \mid \sin~b \mid}{300~{\rm pc}}
+\frac{D x~\cos~b}{3.5~{\rm kpc}}\right)
=\frac{dn_0}{dM}~H_d(x)~, \label{eqno:3a}
\end{equation}
where the galactic longitude $l=0^{\circ}$ has been adopted and
$\frac{dn_0}{dM}=\frac{\rho_d(M)}{M}$ with $\rho_d(M)=0.05 M_{\odot}~
{\rm pc}^{-3}$; $H_d$ can be written as follows
\begin{equation}
H_d(x)=\exp\left(\frac{x D~ (1-11.7
\mid \sin~b \mid)}{3.5~{\rm kpc}}\right)~,
\label{eqno:3b}
\end{equation}
using also the fact that $\cos~b \approx 1$ for the galactic bulge.

For the contribution of the halo dark matter in the form of massive
objects (MHO) located in the disk I use the following distribution
for the number density per unit mass $dn/dM$
\begin{equation}
\frac{dn}{dM}=H_h(x)\frac{dn_0}{dM}=\frac{a^2+R^2_{GC}}
{a^2+R^2_{GC}+D^2x^2-2DR_{GC} x \cos \alpha}~\frac{dn_0}{dM},
\label{eqno:5}
\end{equation}
with $dn_0/dM$ the local dark mass number density.
The local dark mass
density is $\rho_0 \simeq 8 \times 10^{-3} M_{\odot}~{\rm pc}^{-3}$.
It is assumed that $dn/dM$ factorizes in functions
of $M$ (or~$\mu$) and $x$. The galactic core radius is
$a = 5.6$ kpc, whereas
$R_{GC} = 8.5$ kpc is
our distance from the centre of the galaxy,
and $\alpha \simeq~ 4^{\circ}$ is
the angle between the line of sight and the direction of the galactic
centre.

In computing $\Gamma$ one must also take
into account the fact that both the source and
the observer are in motion (Griest 1991; Griest et al. 1991).
Relevant are only the velocities transverse
to the line of sight.
The transverse
velocity of the microlensing tube at position $xD$ is:
$\vec v_t(x)=(1-x)\vec v_{\odot \perp}+x \vec v_{s \perp}$,
and its magnitude is
\begin{equation}
v_t(x)=\sqrt{(1-x)^2 \mid \vec v_{\odot \perp} \mid^2+x^2\mid \vec
v_{s \perp}\mid^2+2x(1-x)\mid \vec v_{\odot \perp}\mid \mid
\vec v_{s \perp}
\mid \cos~\theta }~, \label{eqno:5a}
\end{equation}
where $\vec v_{s \perp}$ and $\vec v_{\odot \perp}$ are
the source and the
solar velocities transverse to the line of sight and $\theta$ the
angle between them.

For the velocity distribution of the MHOs or the faint disk stars I
consider an isothermal spherical model, which in the rest
frame of the galaxy is given by
\begin{equation}
f(\vec v) d^3v=\frac{1}{\tilde v^3_H \pi^{3/2}}~
e^{-\vec v^2/\tilde v^2_H}~d^3v~ . \label{eqno:4}
\end{equation}
Since only the transverse velocities are of relevance cylindrical
coordinates can be used
and the integration made over the velocity component parallel to the
line of sight. Moreover, due to the velocities of the observer and the
source, the value of the transverse velocity gets shifted
by $\vec v_t(x)$.
The distribution for the transverse velocity is thus
\begin{equation}
\tilde f(v_T) dv_T=\frac{1}{\pi v^2_H}~
e^{-(\vec v_T-\vec v_t)^2/v^2_H}~v_T~dv_T~,
\label{eqno:4b}
\end{equation}
where $v_H$ is the velocity dispersion for which I adopt the value
$v_H \approx 30~{\rm km}~{\rm s}^{-1}$ (Paczy\'nski 1991).
The random velocity of the source stars are
again described by an
isothermal spherical distribution, whose transverse
velocity distribution is
\begin{equation}
g(v_{s \perp})dv_{s \perp}=\frac{1}
{\pi v^2_D}~e^{-v^2_{s \perp}/v^2_D}~v_{s \perp}~
dv_{s \perp}~,  \label{eqno:4c}
\end{equation}
where $v_D=156~{\rm km}~{\rm s}^{-1}$
is the velocity dispersion (Mihalas \& Binney 1981; Paczy\'nski 1991;
Griest et al. 1991).

Taking all the above facts into account,
$\Gamma$ turns out to be (
De R\'ujula et al. 1991; Griest 1991)
\begin{eqnarray}
\Gamma &=& 4~r_E D~v_H~u_{TH} \left( \int_0^{\infty}
\sqrt{\mu}~\frac{dn_0}{d\mu}~d\mu
\right) \int_0^{2\pi} d\theta ~\int_0^{\infty} dv_{s \perp}~
g(v_{s \perp}) \nonumber \\ & &
{}~\int_0^1 \sqrt{x(1-x)}~H_i(x)~e^{-\eta^2}
\int_0^{\infty} dy~y^2~I_0(2y\eta)~e^{-y^2}~, \label{eqno:4d}
\end{eqnarray}
where $y=\frac{v_T}{v_H}$, $\eta(x,\theta,v_{s \perp},v_{\odot \perp})=
\frac{v_t}{v_H}$, and $I_0$ is the modified Bessel function of order
0.
In the limit of stationary observer and source star ($v_{\odot \perp}=
v_{s \perp}=0)$,
$\eta=0$ and $I_0=1$; $H_i$ means either $H_d$ or
$H_h$; $u_{TH}$ is related to the minimal experimentally detectable
magnification $A_{TH}=A[u=u_{TH}]$; $v_{\odot \perp}$ is
$\approx 220~{\rm km}~{\rm s}^{-1}$ (more precisely it should
be multiplied by $\cos~ l$,
where $l$ is the galactic longitude
but since $l = 1^{\circ}$, $\cos~ l \approx 1$).
In computing $\Gamma$ one should also take into account the limited
measurable range for the event duration $T$, which translates into a
modification of the integration limits. A fact that can be
described by introducing an efficiency function $\epsilon(\mu)$
(De R\'ujula et al. 1991). However for the range of interest
here, $\epsilon(\mu) \approx 1$, and thus I will neglect it in the
present calculations.

For an experiment monitoring $N_{\star}$ stars during a total
observation
time $t_{obs}$ the number of expected microlensing events is
\begin{equation}
N_{ev}=N_{\star}~t_{obs}~\Gamma ~.
\label{eqno:6}
\end{equation}
Assuming a delta-function-type distribution for the masses
\begin{equation}
\frac{dn_0}{d\mu}=\frac{\rho}{M_{\odot}}~\frac{\delta(\mu-\bar\mu)}
{\mu}~, \label{eqno:4da}
\end{equation}
eq. (\ref{eqno:4d}) can be integrated.
With $N_{\star}=10^{6}$ stars and $t_{obs}=1$ year one gets
\begin{equation}
N_{ev}= 2.08~ \sqrt{\bar\mu}~\left(\frac{\rho_d}{5 \times
10^{-2} M_{\odot}~
{\rm pc}^{-3}} \right)~u_{TH}~, \label{eqno:4e}
\end {equation}
for $H_i=H_d$, and
\begin{equation}
N_{ev}= 0.61~ \sqrt{\bar\mu}~
\left(\frac{\rho_0}{8 \times 10^{-3} M_{\odot}
{}~{\rm pc}^{-3}} \right)~u_{TH}~, \label{eqno:4f}
\end{equation}
for $H_i=H_h$. The numerical factor in
eq. (\ref{eqno:4e}) for
$N_{ev}$ as a function of $b$, the galactic latitude,
varies between 6.8 for $b=0^{\circ}$ and 1.9 for $b=5^{\circ}$,
whereas the factor
for $N_{ev}$ of eq. (\ref{eqno:4f}) remains practically unchanged.\\

\noindent{\bf 3. Most likely mass and mass moments}\\

The probability $P$ that a microlensing event of duration $T$ and
maximum amplification $A_{max}$ be produced by a MHO or a faint disk
star
of mass $\mu$ can be derived starting from eq. (\ref{eqno:4d})
(Jetzer \& Mass\'o 1993) and leads to
\begin{eqnarray}
P(\mu,T) &\propto &\frac{\mu^2}{ T^4} \int_0^{2\pi} d\theta
\int_0^{\infty} dv_{s \perp}~ g(v_{s \perp}) \int_0^1 dx~ (x(1-x))^2~
H_i(x)~e^{-\eta^2}~\nonumber \\ & &
\exp \left(-\frac{r_E^2 \mu x(1-x)}{v_H^2 T^2} \right)
{}~I_0\left(2\eta\frac{r_E \sqrt{\mu x(1-x)}}{v_H T} \right)~.
\label{eqno:8}
\end{eqnarray}
One sees that $P(\mu, T)=P(\mu/ T^2)$, and that it does not depend
on the value of $A_{max}$. In table 1, $P$ is listed for the
four events of OGLE for $H_d$; in table 2 the corresponding
values for the MACHO and EROS events are given, which are computed
using the corresponding formula for $P(\mu,T)$ as discussed in Jetzer
\&
Mass\'o (1994).

The normalization of $P$
is arbitrarily
chosen such that the maximum of $P(\mu_{MP}, T)=1$, and $\mu_{MP}$ is
the most probable value.
The maximum corresponds to
 $\mu r_E^2/v^2_H T^2 \simeq 213$ for $H_d$ and $\simeq 196$ for
 $H_h$.
For the LMC the maximum corresponds to
 $\mu r_E^2/v^2_H T^2=13.0$ (with $v_H = 210~{\rm km}~
{\rm s}^{-1}$ and $r_E = 3.17
\times 10^9$ km).
The 50\% confidence interval
embraces for the mass $\mu$ approximately
the range $1/3\mu_{MP}$ up to $3 \mu_{MP}$.

A more systematic way to extract information on the masses is to use
the
method of moments as discussed in De R\'ujula et al. (1991). The
moments
$<\mu^m>$ are given by
\begin{equation}
<\mu^m>=\int d\mu \frac{dn_0}{d\mu}\mu^m~, \label{eqno:10}
\end{equation}
and $<\mu^m>$ is related to $<\tau^n>=\sum_{events} \tau^n$,
with $\tau \equiv (v_H/r_E) T$, as constructed
from the observations by
\begin{equation}
<\tau^n>=\int dN_{ev} \tau^n=V u_{TH}~\gamma_i(m)  <\mu^m>~,
\label{eqno:11}
\end{equation}
with $m \equiv (n+1)/2$ and
\begin{equation}
V \equiv 4 N_{\star} t_{obs}~ D~ r_E~ v_H~,
\label{eqno:12}
\end{equation}
\begin{equation}
\gamma_i(m) \equiv \int_0^{2\pi} d\theta \int_0^{\infty} dv_{s \perp}~
g(v_{s \perp})
\int_0^1 dx~H_i(x)~(x(1-x))^m~e^{-\eta^2} \int_0^{\infty} dy~
y^{3-2m}~e^{-y^2}~I_0(2\eta y)~.
\label{eqno:13}
\end{equation}
In table 3 the values of $\gamma_i(m)$ are listed for $m$ = 0, 0.5 and
1,
which are the ones needed here.

Eq. (\ref{eqno:13}) is also useful for computing
the average duration $< T >$ (with $T=\frac{R_E}{v_T}=
\frac{r_E}{v_T} \sqrt{\mu x(1-x)}$~) for a microlensing event
defined as
\begin{equation}
< T > = \frac{\int dN_{ev}~\frac{r_E}{v_T} \sqrt{\mu
x(1-x)}}{N_{ev}}~.
\label{eqno:13a}
\end{equation}
Assuming a delta-function-type
mass distribution (eq. (\ref{eqno:4da})),
one finds
\begin{equation}
< T >_i~= \frac{r_E}{v_H}~\sqrt{\bar \mu}~\frac{\gamma_i(1)}
{\gamma_i(0.5)} ~. \label{eqno:13b}
\end{equation}
For disk stars with $H_i=H_d$ one gets
\begin{equation}
< T >_d~\simeq~52 ~\sqrt{\bar\mu} ~~{\rm days}~, \label{eqno:13c}
\end{equation}
and for MHO with $H_i=H_h$
\begin{equation}
< T >_h~\simeq~42~\sqrt{\bar\mu}~~{\rm days}~. \label{eqno:13.d}
\end{equation}

The mean local density of MHO (number per cubic parsec)
is $<\mu^0>$. The average local mass density in MHO is
$<\mu^1>$ solar masses per cubic parsec.
The mean MHO mass, which one gets from
the four events of OGLE, using the faint disk star distribution $H_d$,
is
\begin{equation}
\frac{<\mu^1>}{<\mu^0>} =
\frac{<\tau^1>}{<\tau^{-1}>}~\frac{\gamma_d(0)}
{\gamma_d(1)}~ \simeq 0.28 M_{\odot}.
\label{eqno:aa}
\end{equation}
If instead one uses the $H_h$ distribution, one recovers practically
the
same value for the average, namely: $<\mu^1>/<\mu^0>
\simeq 0.29 M_{\odot}$.
The average value clearly suggests that these objects are faint disk
stars. Notice that $\gamma_i(m)$ depends on the value of the galactic
latitude $b$. However the ratios change much less, in particular
$\gamma_d(0)/\gamma_d(1)$ varies by at most 10 to 15\% by varying $b$
from $0^{\circ}$ to $5^{\circ}$, whereas $\gamma_h(0)/\gamma_h(1)$
remains
practically unchanged. The same being true for the ratios
$\gamma_d(1)/\gamma_d(0.5)$ and $\gamma_h(1)/\gamma_h(0.5)$.

Similarly one can also find the mean MHO mass based on the five
events detected
by EROS and MACHO. The corresponding values for $\gamma(m)$
are $\gamma(0)=0.280$ and $\gamma(1)=0.0362$ (De R\'ujula et al.
1991), thus giving
\begin{equation}
\frac{<\mu^1>}{<\mu^0>} \simeq 0.08 M_{\odot}.
\label{eqno:a1}
\end{equation}
This value is somewhat lower than the one previously computed of
0.14$M_{\odot}$ (Jetzer \& Mass\'o 1994), which was based on only
three events.
The lower value of the MHO mean mass for the halo objects indicates
that they are brown dwarfs rather than low-mass stars as seems
to be the case for the events detected by OGLE.

Although based on few events, these results are of
interest. Clearly, once more data will be available
it will be possible to get more precise information.
In particular, it will be possible to determine other important
quantities
such as the statistical error in eq. (\ref{eqno:aa}) or eq.
(\ref{eqno:a1})
and the fraction
$f \equiv {M_J}/{\rho_0} \sim 0.12~{\rm pc}^3 <\mu^1>$
of the local
dark mass density (the latter one given by $\rho_0$) detected
in the form of MHOs.\\

\noindent{\bf Acknowledgements}\\

I would like to thank the CERN TH Division for hospitality while
this work was carried out.\\

\noindent{\bf References}\\
Alcock, C. et al., 1993, Nature {\bf 365}, 621.\\
Aubourg, E. et al., 1993, Nature {\bf 365}, 623.\\
Bahcall, J.N., \& Soneira, R.M., 1980, Ap. JS. {\bf 44}, 73.\\
De R\'ujula, A., Jetzer, Ph., \& Mass\'o, E., 1991,
MNRAS {\bf 250}, 348.\\
De R\'ujula, A., Jetzer, Ph., \&
Mass\'o, E., 1992, A \& A {\bf 254}, 99.\\
Griest, K., 1991, Ap. J. {\bf 366}, 412.\\
Griest, K. et al., 1991, Ap. J. {\bf 372}, L79.\\
Jetzer, Ph., \& Mass\'o, E., 1994, Phys. Lett. {\bf B323}, 347.\\
Mihalas, D., \& Binney, J., 1981, Galactic Astronomy (San Francisco:
W.H.
Freeman and Co.).\\
Paczy\'nski, B., 1986, Ap. J. {\bf 304}, 1.\\
Paczy\'nski, B., 1991, Ap. J. {\bf 371}, L63.\\
Udalski, A. et al., 1993, Acta Astron. {\bf 43}, 289.\\

\newpage
\vskip 1.0cm
Table 1: Values of the
most probable mass $\mu_{MP}$ in units of $M_{\odot}$ as
obtained by $P(\mu, T)$ with $H_i=H_d$ and $T=\frac{R_E}{v_T}$ for the
four microlensing events of OGLE. ($v_H = 30~{\rm km}~{\rm s}^{-1}$ and
$r_E = 1.25 \times 10^9~{\rm km}$.)

\begin{center}
\begin{tabular}{|c|c|c|c|c|}\hline
  & 1 & 2 & 3 & 4 \\
\hline
$ T$ (days) & 25 & 45 & 10.7 & 14 \\
\hline
$\tau (\equiv \frac{v_H}{r_E} T)$ & $5.2 \times 10^{-2}$ &
$9.35 \times 10^{-2}$ &
$2.22 \times 10^{-2}$ & $2.91 \times 10^{-2}$ \\
\hline
$\mu_{MP}$ & 0.57 & 1.85 & 0.105 & 0.18 \\
\hline
\end{tabular}
\end{center}
\vskip 1.0cm

Table 2: Values of the most probable mass $\mu_{MP}$ in $M_{\odot}$
units for the five microlensing events detected in the LMC ($A_{i}$
= American-Australian
collaboration events ($i$ = 1, 2, 3);
$F_1$ and $F_2$
French collaboration events).
For the LMC: $v_H = 210~{\rm km}~{\rm s}^{-1}$ and
$r_E = 3.17 \times 10^9~{\rm km}$.

\begin{center}
\begin{tabular}{|c|c|c|c|c|c|}\hline
  & $A_{1}$ & $A_{2}$ & $A_3$ & $F_1$ & $F_2$  \\
\hline
$ T$ (days) & 16.9 & 9 & 14 & 27 & 30 \\
\hline
$\tau (\equiv \frac{v_H}{r_E} T)$ & $9.67 \times 10^{-2}$ &
$5.15 \times 10^{-2}$ & $8.01 \times 10^{-2}$ & $1.54 \times 10^{-1}$ &
$1.72 \times 10^{-1}$ \\
\hline
$\mu_{MP}$ & 0.12 & 0.03 & 0.08 & 0.31 & 0.38 \\
\hline
\end{tabular}
\end{center}
\vskip 1.0cm
Table 3: Values of $\gamma_d(m)$ (for $H_d$) and $\gamma_h(m)$
(for $H_h$) for $m$ = 0, 0.5 and 1 valid for $b=-3.^{\circ}9$.

\begin{center}
\begin{tabular}{|c|c|c|}\hline
$m$  & $\gamma_d(m)$ & $\gamma_h(m)$ \\
\hline
0  & 252.7 & 390.3 \\
\hline
0.5 & 14.7 & 27.2  \\
\hline
1 & 1.6  & 2.4  \\
\hline
\end{tabular}
\end{center}

\end{document}